\documentstyle[12pt]{article}
\begin{document}

\baselineskip=24pt


\begin{center}
\large {\bf Gravitational Lensing and the Variability of G}
\end{center}
\begin{center}

{\bf Lawrence M Krauss$^1$ and Martin White$^2$}

Center for Theoretical Physics

Sloane Laboratory, Yale University

New Haven CT 06511
\end{center}

\vskip .3in

\centerline{\bf Abstract}
\noindent The four observables associated with gravitational lensing of
distant quasars by intervening galaxies: image splittings, relative
amplifications, time delays, and optical depths, provide separate measures of
the strength of the gravitational constant $G$ at cosmological distances.
These allow one, in principle, to factor out unknown lensing parameters to
directly to probe the variation of $G$ over cosmological time.  We estimate
constraints on $\dot{G}$ which may be  derivable by this method both now and
in the future.
The limits one may obtain can compete or exceed other direct limits on
$\dot{G}$ today, but unfortunately extracting this information, is not
independent  of the effort to fix other cosmological parameters such as $H_0$
and $\Omega_0$ from lensing observations.


\noindent $|^1$ Also Department of Astronomy.  Research supported in part by a
Presidential Young Investigator Award, the DOE, the Foundation for Physical
Science, and the Texas National Research Laboratory Commission.$|$

\noindent $|^2$ Address after September 1, Center for Particle Astrophysics,
University of California, Berkeley.$|$

\newpage

\noindent {\bf 1. Introduction}

The gravitational constant, $G$, is the poorest measured fundamental constant
in nature.  In fact, it may not even be a constant at all.  The exceedingly
small value of $G$, coupled with the large value of the age of the universe
encouraged speculation early on, first following Dirac and then spurred by the
advent of Brans-Dicke cosmology, that the two quantities may be somehow tied
together \cite{Dirac,Dyson}.
Moreover, because classical general relativity cannot be quantized, there has
been a recurring interest in the possibility that GR arises as the low energy
limit of a more fundamental theory.  In such a theory, the gravitational
constant may arise dynamically, associated with the vacuum expectation value
of some field (or dynamics of some internal space).
Since this dynamical value may be time dependent, so may $G$.  Over the past
year, largely as a result of considerations based on extensions of the original
old inflationary models \cite{Guth,Linde,Albrecht,Steinhardt}, there has been
a renewed interest \cite{La} in the possibility that the gravitational
constant has varied on cosmological timescales.

There exist several sensitive direct probes of a monotonic change in the
gravitational constant during the present epoch, including the use of pulsar
timing measurements and radar experiments, all of which suggest
that $\dot{G}/GH\le 0.4$ today. \cite{Shapiro1,Shapiro2,Helling,Reasen,Damour}
At the opposite extreme, calculations of primordial nucleosynthesis put
indirect limits on $\dot{G}$ during the first
seconds of the big bang expansion from limits on the observed Helium
abundance \cite{Accettakrauss}.
If the variation of $G$ has followed a constant power law in time, the latter
limit $(\dot{G}/GH\le 0.01)$ is stronger than the direct limits on the
variation today .
What has been lacking however is any way to directly probe the value
of $G$ at times between these two epochs.  Since it has even been proposed that
$G$ may oscillate in time \cite{Accetta}, a direct measure of $G$ at
intermediate times would be of great interest.
It is the purpose of this paper to suggest that observations of gravitational
lensing could, in principle, provide such a measure, and to investigate the
realistic limits which it may be possible to obtain.

On first thought it is not clear that lensing can constrain $G$.  While the
bend angle which light rays are subject to is directly related to the strength
of the gravitational constant at the time light rays pass the lensing object,
the quantity which enters into all formulas is the product $GM$, where $M$ is
the mass of the lensing object.  Unless $M$ can be determined independently a
separate extraction of $G$ seems impossible.  However, it is not the actual
bend angle which is directly observed in gravitational lensing.  All lensing
observables depend also (in somewhat different ways) upon the distance of the
lensing galaxy and the quasar as inferred from their redshifts.  The distance
redshift relation depends upon the time-averaged value of $G$, which for
redshifts of $O(1)$ can be a significant fraction of the lifetime of the
universe. Thus for any lensing system a prediction of one observable based on
a measurement of another can give a signal of the time variability of $G$.
What remains to be seen however is exactly how sensitive such a comparison is,
and how much it depends on our knowledge, or lack thereof, of cosmological
parameters such as the Hubble constant $H_0$, the density parameter $\Omega_0$,
and even the cosmological constant $\Lambda$.

The organization of the paper is as follows: in section 2 we outline our
notations and conventions and introduce the models we will use.  In section 3
we consider constraints from lensing statistics and in section 4 we discuss
constraints which can be derived from individual lensing systems.  Section 5
contains our conclusions.

\vskip .2in
\noindent {\bf 2. Cosmology and Lens Models}

The observables of interest in gravitational lensing depend upon the
combination GM (where M is the mass of the lensing galaxy) and the distance to
the galaxy and source.  If it is assumed that the bending occurs predominantly
as the light rays pass through the local region of the lensing galaxy and thus
the time required is much shorter than the time scale
over which G varies significantly, the  effect of the variation of G will
be to replace GM by $G_lM$ (where $G_l$ is  G at the time of lensing) and also
to alter the distance-red shift relation. Measuring the first effect is
cleanest, in principle, because it is not dependent upon cosmological
modelling.  Unfortunately, unless there is an independent way to determine the
mass of the lensing galaxy the first effect alone is unmeasurable.  Since both
velocity dispersion, and stellar luminosity will also depend upon $G$, there
are no observables which seem to allow $M$ to be independently extracted.

Hence, to proceed, we must consider some specific cosmological model,
incorporating a variable $G$. We will consider for definiteness a Brans-Dicke
(BD) theory  (this is perhaps the simplest viable extension of GR with a
varying gravitational constant and is often used in connection with extended
inflationary models).  While our discussion will be in terms of BD cosmology,
the general features should be characteristic of any model with varying $G$. In
particular, these ideas could be applied to any theory based on the
Friedman-Robertson-Walker metric (which has gained more experimental support
recently from the isotropy of the cosmic microwave background) with an
evolution equation for the scale factor determined by the equation of state
of matter which also incorporates a varying value of $G$ consistently in the
equations of motion (e.g. \cite{Dyson,Beckenstein1,Beckenstein2}).

In the BD cosmology the line element is the usual FRW metric
\begin{equation}
ds^2 = -dt^2 + R^2(t) [ d\chi^2 + s_k(\chi)^2 d\Omega ] \end{equation}
where $s_k = \sinh(\chi),\chi,\sin(\chi)$ for $k=-1,0,1$.
Einstein's equations are modified and a new dynamical field $\phi$, with
$G\sim\phi^{-1}$, is introduced.
For large time the general solution of the Brans-Dicke field equations will be
matter dominated and in many cases of interest (e.g.
\cite{Weinberg,La,Steinhardt}) $\phi\sim R^{\sigma}$, where $\sigma$ will be
a function of the Brans-Dicke scalar-tensor coupling constant$^3$,
$\omega$, which tends to zero in the limit $\omega\rightarrow\infty$
(where Einstein's theory is recovered).\newline
$|^3$ The independent limits on the scalar-tensor coupling constant in the
simplest  Brans-Dicke theory are already far more stringent than we will place
from the variation of $G$ \cite{Reasonberg}, but our purpose here is to use
this model merely as a testing ground to explore the sensitivity of lensing
parameters to $G$.$|$ \newline
In the $(k=0)$ examples cited above $\sigma=(1+\omega)^{-1}$. The evolution
equation for the scale factor in a matter dominated epoch is
\begin{equation}
 \left( \dot{R}\over R\right)^2 + {k\over R^2} = {8\pi G_0\over 3}\rho_0
\left({2\omega+3\over 2\omega+4}\right) \left({R_0\over R}\right)^{3+\sigma} +
({\sigma^2\omega\over6}-\sigma)\left( {\dot{R}\over R}\right)^2
\end{equation}
If we define $\eta=R/R_0$ and
\begin{equation}
  \Omega_0=\left( {8\pi G_0\rho_0\over 3H_0^2} \right)
  \left( {2\omega+3\over2\omega+4} \right)
  \left( 1+\sigma-{\sigma^2\omega\over 6} \right)^{-1}
\end{equation}
we can rewrite this as
\begin{equation}
 \dot{\eta}^2 + (\Omega_0-1) H_0^2 = \Omega_0 H_0^2 \eta^{-(1+\sigma)}
\label{eqn:einstein}
\end{equation}
\begin{equation}
  {k\over R_0^2} = \left(\Omega_0-1\right) H_0^2
    \left( 1+\sigma-{\omega\sigma^2\over 6} \right)
\label{eqn:konR2}
\end{equation}

Note that if one instead were simply to allow $G$ to vary as a power law and
were to use Einstein's equations unchanged one would obtain the same result
(\ref{eqn:einstein},\ref{eqn:konR2}), except without the last factor on the
r.h.s. of (\ref{eqn:konR2}).  For a fixed value of $\Omega_0$ and $H_0$, it is
this factor which causes $R_0$ to vary with $\omega$.

The measure of distance we will use is the angular diameter distance
\begin{equation}
d_A   =  {R_0\over 1+z} s_k(\chi)
\end{equation}
which assumes that the lensed rays traverse a mean filled ``beam".
(Similar, but algebraically more complex constraints can be obtained in the
case of an ``empty" beam approximation, where affine angles and distances are
used (i.e. see \cite{Turner,White}). The mean filled beam approximation is
probably closer to the actual situation, however, it has been shown that the
uncertainty due to clumpiness of matter in the beam trajectory can be one of
the main sources of uncertainty in the analysis of individual lensing systems
\cite{Alcock}.  If the Hubble constant were independently measured it is
possible that this uncertainty could be reduced since it also enters into the
determinations of $H_0$ from time delays in lensing systems.

Once the evolution equation for the scale factor, $R(t)$, is specified we can
solve for the distance redshift relation in the usual way \cite{Weinberg}.
For the special case $k=0$ we obtain a simple expression for the distance as
a function of $x=1+z$, (similar expressions for $\Omega_0 \not= 1$ can also be
obtained.)
\begin{equation}
d_A  = {1\over\beta H_0 x} (1-x^{-\beta})
\end{equation}
where $\beta=(1+\sigma)/2$.  The distance is plotted in figure 1 for
$\beta=0.4,0.5,0.6$, and reduces to the usual expression \cite{Turner} in the
limit of constant G ($\omega\rightarrow\infty, \beta\rightarrow 1/2$).
As a guide to the expected magnitude of $\beta$ one would like to obtain
sensitivity to, notice that an assumed variation-since-lensing of
\begin{equation}
{\Delta G\over G} = {G_l-G_0\over G_0} = 20\% \Rightarrow
\sigma = {\log 1.2\over \log x_l} \sim 0.2
\end{equation}
for lenses $z_l\sim 1.5$.  This corresponds to $\beta\sim 0.6$.  If we take
the age of the universe to be $t_0 = 2/(3+\sigma) H_0^{-1} = 10^{10}$yr, this
then gives
\begin{equation}
\dot{G}/G |_0 = -\sigma H_0 \sim 10^{-11} \mbox{yr}^{-1}
\end{equation}
which is comparable with other direct measures of $\dot{G}/G$.

We will consider two simplified lens models in what follows:
the point mass and the isothermal sphere lenses.  The point mass lens
is chosen for its simplicity, the isothermal lens because the flatness of
rotation curves of galaxies suggest $\rho\sim r^{-2}$ is a reasonable
approximation to galactic mass distributions (at least asymptotically).
While for any actual lens system these models are overly simplistic they
serve to illustrate the main points. The observables for these lens systems
which we would want to examine for sensitivity to $\beta$ are:
time delays between images, angles between images and ratio in brightness of
the images, as well as the (differential and total) optical depth for lensing.

\clearpage

\noindent {\bf 3. Lensing statistics}

The formalism appropriate to gravitational lensing statistics was first
developed in \cite{Turner} and latter generalized to arbitrary
Robertson-Walker cosmologies in \cite{Gott} (see also \cite{White} for
a recent presentation).  The key quantity is the optical  depth, or integrated
probability of lensing, $\tau$, assuming a non-evolving  population of
galaxies, modelled as singular isothermal spheres.
This depth is relatively free of matter clustering uncertainties \cite{Alcock}
but is sensitive to variations in the distance-redshift relation, which makes
it a good probe of cosmology.

As an example consider the expression for $\tau$, for a $k=0$ universe.
Including the $\beta$ dependence of the bend angle $(\alpha\sim G)$ we obtain
\begin{equation}
\tau\left(y=1+z\right) = {F\over\beta^2}\int_1^y dx\ x^{-3-\beta}
(x^{\beta}-1)^2 \left[{ y^{\beta}-x^{\beta} \over y^{\beta}-1 }\right]^2
\ \ \ ;\  F = {n_0\pi\alpha_0^2 R_0^3}
\label{eqn:tau}
\end{equation}
where $H_0 R_0 = 1$ and $n_0$ is the comoving number density of galactic
lenses, which we assume in this instance are all identical (non-evolving)
isothermal spheres producing identical bend angles $\alpha_0$.  The bend
angle $\alpha_0$ is related to measured velocity dispersions of nearby
galaxies today.  For further details of these definitions see \cite{Turner}.
Equation (\ref{eqn:tau}) reduces to eqn (2.26c) of \cite{Turner} in the limit
$\beta\rightarrow 1/2$.
The integral can be done analytically but the result is cumbersome and is not
shown here.   The optical depth vs redshift is shown in figure 2 for
$\beta=0.4,0.5,0.6$. As can be seen the variation with $\beta$ is slight
making this a poor measure of $\dot{G}/G$.
A similarly small dependence on $\beta$ is shown by the differential optical
depth $d\tau/dz$. Since the distance-redshift relation becomes less $\beta$
dependent as  $\Omega_0$ decreases we expect the variation in $\tau$ to be
less than above when $\Omega_0<1$, although this is somewhat offset by the
$\beta$ dependence of $F$ coming through $R_0$. Thus variations of
$G$ going as a power law in time have little effect on lensing statistics, at
least at the level where these statistics are likely to be determined in the
forseeable future.

\vskip .2in
\noindent {\bf 4. Individual systems}

A better hope of constraining $\dot{G}$ comes from examining the observables
associated with multiply imaged quasars (i.e. see\cite{Hewitt}).
Specifically we will be interested in the observables: time delay, image
splitting and image magnification for our two model lenses.  The strategy will
be the following:  each observable will depend both on $GM$, and on $d_A (G)$.
If we have more than two observables for each system, then we hope to overly
constrain the system so that we can check for consistency between the
different determinations of these quantities from each observable.

\vskip .1in

\noindent 1) Point mass lens.

For these lenses the time delay $\Delta t$ and ratio in magnitude of images $r$
are related through \cite{Krauss}
\begin{equation}
\Delta t = 2GM (1+z)[ (r-{1\over\sqrt{r}}) + \log(r)]
\end{equation}
(the $(1+z)$ factor is absent in microlensing \cite{Krauss}) so these two
parameters can be used to infer $GM$ at the time of lensing, {\it independently
of the cosmological distance-redshift relation}. If it is not possible to
measure $\Delta t$ in the lens system, or if the measure has a large
uncertainty, $GM$ must be obtained some other way, e.g. from virial velocity
measurements.
Any limit on $\beta$ will depend on how well this quantity is known.

Given $GM$ we can use the observed angular splitting of images and the relation
\begin{equation}
\Delta\theta = {4GM\over c^2 S} {r-1\over\sqrt{r^{1/2}(r-1)-2r}}
\end{equation}
to determine $S=D_S/D_{LS}$ where $D_S$ and $D_{LS}$ are the angular diameter
distances from the observer to the source and from the lens to the source
respectively.  This is a function only of the (known) redshifts, $\Omega_0$
and $\beta$, e.g. in the $k=0$ case
\begin{equation}
  S \equiv  {s_0(\chi_S)\over s_0(\chi_S-\chi_L)} =
  {1-x_S^{-\beta}\over x_L^{-\beta}-x_S^{-\beta}}
\end{equation}
so a knowledge of the redshifts allows a determination of $\beta$ (up to
clumping uncertainties \cite{Alcock}) if we assume a value for $\Omega_0$
(or conversely a determination of $\Omega_0$ if we know $\beta$).
Notice that S is a ratio of distances and so is independent of $H_0$.
As an example if we take $z_L=1, z_S=3$ then for $k=0$, mean filled beam,
S is a monotonically increasing function of $\beta$ varying from $2.3$ to
$2.5$ as $\beta$ runs from $0.4$ to $0.6$ as can be seen in figure 3.
Given the above and the fact that typical image splittings can be
$\sim 3''-7''$ it is not impossible that a good measurement of the
angular splitting could limit $\beta$ to be in the range competitive with other
direct probes of $\dot{G}$.

\vskip .1in

\noindent 2) Isothermal Sphere

For the somewhat more realistic, isothermal sphere model the situation is
simpler (in principle).
If the velocity dispersion of the lensing galaxy, $\sigma_{||}$, is known, say
from measurements of the rotation curves, a measure of the angular splitting
allows us to immediately infer S:
\begin{equation}
  \Delta\theta = {2\alpha\over c^2 S} = {8\pi\sigma_{||}^2\over c^2 S}
  \Rightarrow
  S = {8\pi\over\Delta\theta}\left( {\sigma_{||}\over c} \right)^2
\end{equation}
In fact a simultaneous measurement of the time delay, which gives us $D=D_L/S$
(where $D_L$ is the angular diameter distance from the observer to the lens),
\begin{equation}
  \Delta t = 32\pi^2(1+z)\left({\sigma_{||}\over c}\right)^4 {D\over c}
\end{equation}
and the image splitting would in principle allow us to measure both $\beta$
and $\Omega_0$ (up to uncertainties in $H_0$, because $D_L$ has the dimensions
of distance and hence is dependent upon $H_0$)
because the dependence of $D$ and $S$ on $\beta$ and $\Omega_0$ is
different$^4$.
We expect the strongest constraint on $\beta$ for fixed $\Omega_0$ to come
from $S$ however. \newline
$|^4$ See figure 4.$|$ \newline

The singular isothermal sphere (SIS) model is probably still
too naive to apply to actual individual lens systems.  One should at least
include the effects of a finite galactic core \cite{Hinshaw,White}.
Alternatively, a more complicated, but more general model, the `elliptical
lens'\newline
\cite{Narayan}, is available for use in extracting these quantities
for individual galactic lenses.  Nevertheless, the SIS model should give a
general idea of the methodology to be used, and the possible sensitivity to
$\beta$.  Using these other models in an application of these ideas to actual
lenses would merely require replacing the above equations for $S$ and
$\Delta t$ with somewhat more complicated equations which would include the
lensing parameters fit by the observations.

Because of the simplicity of the SIS model we only had to make due with 2
lensing observables to overconstrain the system.  We note however that the
ratio of image amplifications itself is also dependent on $S$ and $\beta$,
and so can also be used to probe for consistency when more complicated fits
to galactic lenses are required.

\vskip .2in

\noindent {\bf 5. Conclusions}

While we have demonstrated here that gravitational lensing provides in
principle a direct sensitivity to variations in $G$ over cosmological time,
our results suggest that to be competitive with limits on $\dot{G}$ at the
present time, lensing parameters must be extracted from observations at
the level of $10\%$ or better--a daunting but not impossible task.
Statistical measures such as the optical depth do not seem sufficiently
sensitive to $\dot{G}/G$, the effects of reasonable changes in $G$ being
swamped by larger uncertainties from our present lack of knowledge of
$\Omega_0$ and $H_0$.  For individual lens systems a simplified model suggests
that it may be possible to see variations of the order
$\dot{G}/G\sim 10^{-11}/$yr if accurate measures of the angular
splitting, amplifications, and perhaps also time delays become available
for a system with source and lens at relatively high redshifts ($z_S\sim 2,3$
and $z_L\sim 1$).
Pessimistically, it is worth noting that a possible variation in $G$ is yet
one more uncertainty which could limit one's ability to extract $H_0$ from
measurements of time delays in individual systems.  On the other hand, if
$H_0$ and  $\Omega_0$ are measured reliably by independent means, one's ability
to probe for variations in $G$ will improve.

Nevertheless, in spite of the limitations of this method, it is worth
emphasizing that it does provide perhaps the only `direct' probe of variations
in $G$ during intermediate times between the present epoch, and the
nucleosynthesis era in the very early universe.  We have placed `direct' in
quotation marks because as we have demonstrated, one's ability to extract
information on $\dot{G}$ is intertwined with our knowledge (or ignorance) of
the proper cosmological model for the evolution of the universe during this
time.  In this regard, we also note that for our analysis, we
used as an analytic tool to probe the sensitivity of lensing, a simple
Brans-Dicke cosmological model, which in fact is already ruled out by other
constraints for the parameter range which would produce the level of time
variation probed here.  In this case, $G$ would vary as a simple power law
with time.  We expect our results would be applicable for any similar, more
viable, model.  Of course, there are other possibilities,
including an oscillatory behaviour of $G$ with a cosmologically interesting
period (i.e. \cite{Accetta}).  One would need specific models to perform an
analysis similar to that performed here, but it may be that for such scenarios,
gravitational lensing could provide sensitive limits, by comparing results
obtained from lensing systems at different redshifts.

\newpage

\parindent .1cm

\noindent {\bf \large References}

Accetta, F.S., Krauss, L.M. and Romanelli, P., 1990, Phys. Lett. { 248}, 146

Accetta, F.S., and Steinhardt, P.J., UPR-0454T

Albrecht, A. and Steinhardt, P.J., 1982, Phys. Rev. Lett. { 48}, 1220

Alcock and Anderson, 1985, Ap.J. { 291}, L29

Beckenstien, J.D., 1977, Phys. Rev. { D15}, 1458;

Beckenstein, J.D., and Meisels, A., 1980, Ap.J. { 237}, 342

Damour, T., Gibbons, G.W., and Taylor, J.H., 1988, Phys. Rev. Lett.
{ 61}, 1151

Dirac, P.A.M., 1937, Nature { 139}, 323;

Dirac, P.A.M., 1938, Proc.Roy.Soc. { A165}, 199

Dyson, F.J., in ``Aspects of Quantum Theory", ed. A.Salam and E.P.Wigner, p213.

Gott R., et. al., 1989, Ap.J. { 338}, 1

Guth, A.H., 1981, Phys. Rev. { D23}, 347

Helling, 1987, in ``Problems in Gravitation", Moscow State U.P. p.46,
\newline\indent\hspace{.1in} ed. V.N.Melnikov

Hewitt, J.N., et. al., 1988, in ``Gravitational Lenses", Springer-Verlag,
\newline\indent\hspace{.1in} ed. Moran, Hewitt and Lo.

Hinshaw, G., and Krauss, L.M., 1987, Ap.J. { 320}, 468

Krauss, L.M., and Small, T.A., 1991, Ap. J. { 378}, 22

Krauss, L.M., and White, M., 1991, Yale preprint YCTP-P42-91,
\newline\indent\hspace{.1in} submitted to Ap. J.

La, P. and Steinhardt, P.J., 1989, Phys. Rev. Lett. { 62}, 376

Linde, A.D., 1982, Phys. Lett. { 108B}, 389;

Narayan, R. and Grossman, S. in ``Gravitational Lenses", ed. J.M. Moran,
\newline\indent\hspace{.1in} J.N. Hewitt and K.Y. Lo, Springer-Verlag (1988)

Press and Gunn, 1973, Ap.J. { 185}, 397

Reasenberg, R.D., et. al., 1979, Ap.J. { 234}, L219

Reasenberg, R.D., 1983, Philos. Trans. Roy. Soc. London { A 310}, 227

Shapiro, I.I., 1964, Phys. Rev. Lett. { 13}, 789;

Shapiro, I.I. et. al., 1971, Phys. Rev. Lett. { 26}, 1132

Steinhardt, P.J. and Accetta, F.S., 1990, Phys. Rev. Lett. { 64}, 2740

Turner, E.,  Ostriker, J., and Gott, R., 1984, Ap.J. { 284}, 1

Weinberg, S., 1972, ``Gravitation and Cosmology", Wiley

Will, C.M., ``Theory and Experiment in Gravitational Physics",

Will, C.M., 1984, Phys. Rep. { 113}, 345

Young et. al., 1980, Ap.J. { 241},  507

\end{document}